\documentclass[pra,reprint,showpacs,amsmath,amssymb,superscriptaddress]{revtex4-1}
\usepackage{graphicx}
\usepackage{color}
\usepackage[colorlinks,bookmarks=false,citecolor=blue,linkcolor=red,urlcolor=blue]{hyperref}
\usepackage{multirow}
\newcommand{\ket}[1]{\mbox{$ | #1 \rangle $}}
\newcommand{\bra}[1]{\mbox{$ \langle #1 | $}}

\begin{document}
\title{Two-photon Rabi-Hubbard and Jaynes-Cummings-Hubbard models: photon pair 
superradiance, Mott insulator and normal phases}
\author{Shifeng Cui}
\affiliation{Department of Physics, Beijing Normal University, Beijing
  100875, China} 
\author{F. H\'ebert}
\affiliation{Universit\'e C\^ote d'Azur, CNRS, INPHYNI, France}
\author{B. Gr\'emaud}
\affiliation{Aix Marseille Univ, Universit\'e de Toulon, CNRS, CPT, Marseille, France} 
\affiliation{MajuLab, CNRS-UCA-SU-NUS-NTU International Joint Research
  Unit, 117542 Singapore}
\affiliation{Centre for Quantum Technologies, National University of
  Singapore, 2 Science Drive 3, 117542 Singapore}
\author{V.$\,$G.  Rousseau}
\affiliation{5933 Laurel St, New Orleans, LA70115, USA}
\author{Wenan Guo}
  \email{waguo@bnu.edu.cn}
\affiliation{Department of Physics, Beijing Normal University, Beijing
  100875, China}
\affiliation{Beijing Computational Science Research Center, Beijing 100193, China}
\author{G. G. Batrouni}
\email{george.batrouni@inphyni.cnrs.fr}
\affiliation{Universit\'e C\^ote d'Azur, CNRS, INPHYNI, France}
\affiliation{MajuLab, CNRS-UCA-SU-NUS-NTU International Joint Research
  Unit, 117542 Singapore}
\affiliation{Centre for Quantum Technologies, National University of
  Singapore, 2 Science Drive 3, 117542 Singapore} 
\affiliation{Department of Physics, National University of Singapore, 2
  Science Drive 3, 117542 Singapore}
\affiliation{Beijing Computational Science Research Center, Beijing 100193, China}

\begin{abstract}
We study the ground state phase diagrams of two-photon Dicke, the
one-dimensional Jaynes-Cummings-Hubbard (JCH), and Rabi-Hubbard (RH)
models using mean field, perturbation, quantum Monte Carlo (QMC), and
density matrix renormalization group (DMRG) methods. We first compare
mean field predictions for the phase diagram of the Dicke model with
exact QMC results and find excellent agreement. The phase diagram of
the JCH model is then shown to exhibit a single Mott insulator lobe
with two excitons per site, a superfluid (SF, superradiant) phase and
a large region of instability where the Hamiltonian becomes
unbounded. Unlike the one-photon model, there are no higher Mott
lobes. Also unlike the one-photon case, the SF phases above and below
the Mott are surprisingly different: Below the Mott, the SF is that of
photon {\it pairs} as opposed to above the Mott where it is SF of
simple photons.  The mean field phase diagram of the RH model predicts
a transition from a normal to a superradiant phase but none is found
with QMC.
\end{abstract}

\pacs{
05.30.Jp
05.30.Rt
42.50.Pq
}

\maketitle

\section{Introduction}
Continued progress in controlling and tuning interactions between
photons and individual atoms has made possible the realization of
elementary quantum electrodynamics building blocks\cite{schoelkopf08}
using two-level atoms in cavities\cite{haroche06} or Josephson
junctions on solid state
chips\cite{wallraff04,chen07,lambert09,nataf10,viehmann11} (circuit
QED).  Generally speaking, such couplings of photons to two-level
systems are well described by the Rabi model\cite{rabi36,rabi37}. In
what is referred to as the strong coupling limit, $g/\omega \lesssim
0.1$, (see below) one can apply the random wave approximation (RWA)
and ignore ``counter rotating'' (CR) terms which do not conserve the
number of excitons (the number of photons plus the number of excited
atoms). When these terms are ignored, we obtain the Jaynes-Cummings
model\cite{jaynes63} which, due its $U(1)$ symmetry, conserves the
number of excitons and can be solved exactly for a single cavity. An
ensemble of such cavities can be connected (here we consider a
one-dimensional chain) by tuning the tunneling rate of the photon
modes between near neighbor cavities resulting in a lattice model, the
Jaynes-Cummings-Hubbard model, which consists of itinerant bosons
(photons) hopping between near neighbor sites and interacting with
localized two-level atoms (``spins'' or qubits). The properties of
this model were shown to be very similar to those of the
one-dimensional Bose-Hubbard model (BHM)\cite{batrouni90} exhibiting a
superfluid phase (of excitons) and incompressible Mott insulator (MI)
lobes\cite{greentree06,hartmann06,rossini07,koch08,hohenadler08,hartmann08,schmidt09,zhao08}.
The MI is essentially a product single site states composed of a
superposition of photons and excited atoms\cite{schmidt09} and
exhibits behavior similar to photon blockade\cite{birnbaum05} where
there is a finite energy gap opposing the addition of a photon. Long
range frustrated hopping\cite{hohenadler12} and dynamic properties of
this model have also elicited much
interest\cite{pippan09,tomadin10,tomadin10b,hohenadler11,carusotto13}. When the coupling
becomes of the order of the photon frequency, $g/\omega$, the
``ultra-strong coupling'' regime which has been achieved
experimentally\cite{niemczyk10,forndiaz10,chen17,forndiaz17,
  yoshihara17}, the RWA is no longer valid and the counter rotating
terms come into play reducing the symmetry from $U(1)$ to $Z_2$.  This
has several consequences: (a) CR terms cause the exciton number not to
be conserved thus excluding the possibility of any Mott phases, (b)
the system is now in the universality class of the Ising model with
two phases, a disordered and an ordered (coherent) phase, (c) due to
the discrete nature of the symmetry, it can break spontaneously in one
dimension. Consequently, the ordered coherent phase is a photon
Bose-Einstein condensate (BEC) and the transition between this phase
and the disordered phase exhibits Ising critical
exponents\cite{schiro13,zheng11,schiro13,kumar13,flottat16}. This
transition resembles the incoherent/coherent (normal/superradiant)
phase transition in the Dicke model\cite{hepp73,rotondo15}.

Multi-photon processes have also attracted scrutiny as more
experimental systems are being realized where such physics enters into
play. For example, the two-photon Rabi model has been used as an
effective model to describe second order processes in Rydberg atoms in
cavities\cite{bertet02} and quantum dots\cite{stufler06,delvalle10},
and mechanisms have been proposed to realize it in circuit
QED\cite{felicetti18}. Unlike the one-photon case, the two-photon Rabi
model undergoes spectral collapse, where the Hamiltonian is no longer
bounded from below, when the coupling exceeds a certain
value\cite{emary02,dolya09,travenec12,maciejewski15,travenec15,peng17,chen12,
  felicetti15}. As in the one-photon case, in the strong coupling
limit where the CR terms can be ignored, the Hamiltonian has $U(1)$
symmetry. In the ultra strong coupling limit, where the CR terms must
be restored, the symmetry is reduced to $Z_{2n}$ for the $n$-photon
case, {\it i.e.} $Z_4$ in the two-photon case. The ground state of the
many-body two-photon model was studied in the context of the Dicke
model\cite{dicke54} using mean field methods\cite{garbe17}. A quantum
phase transition between a normal (disordered) and superradiant phase
was predicted. In this model, however, the term superradiant is used
to indicate a macroscopic change in the average number of photons but
which remains relatively small. This is to be contrasted with the
one-photon case where there is a very large number of photons in the
superradiant phase and they form a
BEC\cite{schiro13,zheng11,schiro13,kumar13,flottat16}. The spectral
collapse region was also found.

In this paper we will first use exact QMC simulations to determine the phase 
diagram of the two-photon Dicke model and compare with mean field 
results\cite{garbe17}. This also serves to verify our QMC method. We then 
use our QMC algorithm, and also DMRG calculations, to study the phase diagram 
of another many-body system, namely the one-dimensional two-photon 
Jaynes-Cummings-Hubbard (JCH) model. The numerical results are compared with 
perturbation and mean field calculations. Finally, we use our numerical 
methods to study the full Rabi-Hubbard model. 

Our results below show that the mean field phase diagram of the Dicke model is 
in agreement with our exact QMC simulations. In addition, we find that, 
surprisingly, the Jaynes-Cummings-Hubbard model exhibits only one Mott 
insulating lobe before the systems becomes unstable. Furthermore, doping above 
the MI yields a photon superfluid phase, while doping below the MI yields an 
unexpected photon {\it pair} superfluid phase. QMC simulations suggest that the 
Rabi-Hubbard model does not exhibit a phase transition from a normal to a 
superradiant phase.

The paper is organized as follows. In section II we describe the models and the 
physical quantities we will study. In section III we review the mean field 
calculation for the Dicke model and discuss our exact QMC results. In section 
IV we show our perturbation and exact QMC results for the JCH model followed by 
section V where we discuss the RH model. We present some conclusions in section 
VI followed by appendices A, B, and C where we show details of some of our 
calculations. 

\section{Models}

We will first study the two-photon Dicke model where $N$ two-level systems 
(qubits) couple to a single photon mode and which is governed by the 
Hamiltonian
\begin{eqnarray}
\nonumber
H_D&=& \omega {\hat a}^\dagger {\hat a}^{\phantom\dagger} + 
\frac{\omega_q}{2}\sum_{j=1}^N \sigma^z_j + \frac{g}{N}\sum_{j=1}^N \sigma^x_j 
({\hat a}^2+{\hat a}^{\dagger 2})\\
\nonumber
&=&\omega {\hat a}^\dagger {\hat a}^{\phantom\dagger} + 
\frac{\omega_q}{2}\sum_{j=1}^N \sigma^z_j \\
&&+ \frac{g}{N}\sum_{j=1}^N 
(\sigma^+_j + \sigma^-_j)({\hat a}^2+{\hat a}^{\dagger 2}).
\label{dicke}
\end{eqnarray}
This model was examined with mean field in Ref. \onlinecite{garbe17}. Here,
$\omega$ is the photon frequency, $\omega_q$ the qubit energy spacing, $g$ the 
coupling constant, $\hat a$ (${\hat a}^\dagger$) is the photon destruction 
(creation) operator, $\sigma^z_i$ and $\sigma^x_j=\sigma^+_j+\sigma^-_j$ are 
Pauli matrices acting on the $j$th qubit, $\sigma^+_j$ ($\sigma^-_j$) is the 
corresponding raising (lowering) operator. A related model is the Rabi-Hubbard 
(RH) model,
\begin{eqnarray}
\nonumber
H_{RH}&=& -J \sum_{i=1}^N \left ( {\hat a}^\dagger_i {\hat 
a}^{\phantom\dagger}_{i+1} + h.c. \right ) + 
\sum_{i=1}^N
\left (\omega {\hat a}^\dagger_i {\hat a}^{\phantom\dagger}_i + 
\omega_q \sigma^+_i \sigma^-_i  \right )\\
&&+ g\sum_{i=1}^N \left ( \sigma^+_i  + \sigma^-_i \right ) \left ({\hat 
a}^2_i+{\hat a}^{\dagger 2}_i \right ),
\label{rabihub}
\end{eqnarray}
where now $N$ is the number of sites (or cavities) and ${\hat a}_i$ is
the photon mode in the $i$th cavity. Note that photon modes can tunnel
between sites. Ignoring in Eq.~(\ref{rabihub}) the CR terms,
$(\sigma^+_i{\hat a}^{\dagger 2}_i + \sigma^-_i {\hat a}^2_i$), yields
the Jaynes-Cummings-Hubbard (JCH) model in which the number of
excitons is conserved.

In both models, Eq.~(\ref{dicke}) and Eq.~(\ref{rabihub}), when the CR
terms are dropped, the system is invariant under the generalized
rotation operator,
\begin{equation}
 {\cal R}(\theta) = {\rm exp} \left (i \theta {\hat a}^\dagger_j 
{\hat a}^{\phantom\dagger}_j 
+ i2 \theta \sigma^+_j \sigma^-_j \right ),
\end{equation}
with ${\cal R}(\theta)^\dagger {\hat a}_j {\cal
  R}(\theta)^{\phantom\dagger}={\rm e}^{i\theta} {\hat a}_j$, and
${\cal R}(\theta)^\dagger \sigma^-_j {\cal
  R}(\theta)^{\phantom\dagger}={\rm e}^{i2\theta} \sigma^-_j$ for any
$\theta$, thus exhibiting $U(1)$ symmetry. However, the action of
${\cal R}(\theta)$ on the CR terms introduces a phase ${\rm
  exp}(i4\theta)$ thus reducing the symmetry to $Z_4$, in other words,
the system is left invariant by the rotation only for
$\theta=n2\pi/4$, $n=0,1,2,3$. Therefore, when the CR terms are
ignored, the $U(1)$ symmetry results in the conservation of the number
of excitons, $N_{exc} = N_{photon} + 2N_{+}$ where $N_+$ is the number
of qubits in the excited state. De-exciting a qubit generates two
photons and vice versa. On the other hand, in the presence of the CR
terms, $N_{exc}$ is not conserved, and quantum phase transitions would
be expected to reflect the discrete $Z_4$ symmetry.  Note that this
discrete symmetry can break spontaneously for a one-dimensional
quantum system in its ground state.

To characterize the various possible phases, we calculate several Green 
functions,
\begin{equation}
 G_{\alpha,\beta}(r) \equiv \frac{1}{2N} \sum_i \langle \alpha_i \beta_{i+r} + 
{\rm h.c.} \rangle,
\label{greens}
\end{equation}
where $\alpha$ and $\beta$ denote creation and annihilation operators
of the photons (${\hat a}^\dagger_i$ and ${\hat
  a}^{\phantom\dagger}_i$) or the qubits ($\sigma^+_j$ and
$\sigma^-_j$). For example, the photon Green function at equal time is
given by,
\begin{equation}
 G_{a^\dagger, a^{\phantom\dagger}}(r) = \frac{1}{2N} \sum_i \langle 
{\hat a}^\dagger_i {\hat a}^{\phantom\dagger}_{i+r} + {\hat a}^\dagger_{i+r} 
{\hat a}^{\phantom\dagger}_{i} \rangle.
\label{photongreen}
\end{equation}
The qubit Green function is,
\begin{equation}
 G_{\sigma^-, \sigma^+}(r)=\frac{1}{2N}\sum_i \langle \sigma^-_i
 \sigma^+_{i+r} + \sigma^-_{i+r}\sigma^+_i \rangle,
\end{equation}
and the following two functions will be particularly useful:
\begin{equation}
  G_{\sigma^+, a}(r)=\frac{1}{2N}\sum_i \langle {\hat a}^{\phantom\dagger}_i 
\sigma^+_{i+r} 
+ \sigma^-_{i+r}{\hat a}^\dagger_i \rangle,
\end{equation}
and
\begin{equation}
  G_{\sigma^+, a^2}(r)=\frac{1}{2N}\sum_i \langle {\hat a}^{2}_i \sigma^+_{i+r} 
+ \sigma^-_{i+r}{\hat a}^{\dagger 2}_i \rangle.
\end{equation}
Power law decay of one of these Green functions would indicate quasi-long range 
order for the corresponding quantity. We also measure the average number of 
excitons,
\begin{equation}
 N_{exc} = \sum_i \langle {\hat a}^\dagger_i{\hat a}^{\phantom\dagger}_i + 
2\sigma_i^+\sigma_i^- \rangle,
\label{nexc}
\end{equation}
and the superfluid density,
\begin{equation}
 \rho_s = \frac{\langle W^2\rangle}{2t\beta L^{d-2}},
 \label{rhos}
\end{equation}
where $W$ is the winding number of an exciton and $d$ the dimensionality. 
To study the phase diagrams of these models, we use several methods:
Mean field, perturbation expansion, QMC and DMRG.

\section{The Dicke Model}

First, we will review briefly the mean field results of
Ref.~\onlinecite{garbe17} and then discuss our QMC results. First the
total angular momentum operators are defined, ${\hat J}_z\equiv \sum_i
\sigma^z_i/2$ and ${\hat J}^{\pm}\equiv \sum_i \sigma^{\pm}_i$. Then
using the Holstein-Primakoff (HP) transformation, they define the
bosonic operators:
\begin{equation}
 {\hat J}^+ = {\hat b}^\dagger\sqrt{N-{\hat b}^\dagger {\hat 
b}^{\phantom\dagger}},\,\,\, {\hat J}^- = \sqrt{N-{\hat b}^\dagger 
{\hat b}^{\phantom\dagger}}\, {\hat b}^{\phantom\dagger},\,\,\, {\hat 
J}_z={\hat b}^\dagger {\hat b}^{\phantom\dagger} - \frac{N}{2},
\end{equation}
where $[{\hat b},{\hat b}^\dagger]=1$. Then the bosonic operators are
replaced by the average value in the ground state, ${\hat b}\to {\bar
  b}=\langle GS|{\hat b}|GS\rangle$. Making the substitutions in the
Hamiltonian, Eq.~(\ref{dicke}), yields,
\begin{equation}
 H^{MF}_D = \omega {\hat a}^\dagger{\hat a} + g_{\bar b} ({\hat a}^2+{\hat 
a}^{\dagger 2}) + \omega_q|{\bar b}|^2-\frac{\omega_qN}{2},
\end{equation}
with
\begin{equation}
 g_{\bar b} = \frac{g}{N}({\bar b} + {\bar b}^*)\sqrt{N-|{\bar b}|^2}.
\end{equation}
The Hamiltonian is now a quadratic in the photon operators and can be
diagonalized exactly giving the ground state energy, $E_G$, which is
to be minimized with respect to ${\bar b}$. This determines the value
of the order parameter, ${\bar b}$ as a function of the other
parameters $\omega$, $\omega_q$ and $g$. It was found that for
$g<g_c=\sqrt{\omega\omega_qN/4}$, $E_G$ is minimum for ${\bar
  b}={\bar b}^*=0$. For $g>g_c$, the ground state is twofold
degenerate and the order parameter, acquires a nonzero value, ${\bar
  b}={\bar b}^*$:
\begin{eqnarray}
\label{mftorder}
 {\bar b} &=& \pm \sqrt{\frac{N}{2}}\left ( 1- 
\sqrt{\frac{1-\mu}{4\mu^2\lambda^2-\mu}}  \right )^{1/2}\\
\label{lambdamft}
\lambda &=& \frac{\omega}{2\omega_qN}\geq 0\\
\label{mumft}
\mu&=& \frac{4g^2}{\omega^2}.
\end{eqnarray}
Therefore, the system exhibits two phases, a normal phase (${\bar b}=0$) and a 
superradiant phase (${\bar b}\neq 0$) separated by the transition line 
$g_c=\sqrt{\omega\omega_qN/4}$. Furthermore, for $g>\omega/2$, the 
argument of the square root in Eq.~(\ref{mftorder}) becomes negative indicating 
that the Hamiltonian is unbounded. These results map out the phase diagram as 
shown in Ref.~\onlinecite{garbe17}.

\begin{figure}[!h]
\includegraphics[width=1 \columnwidth]{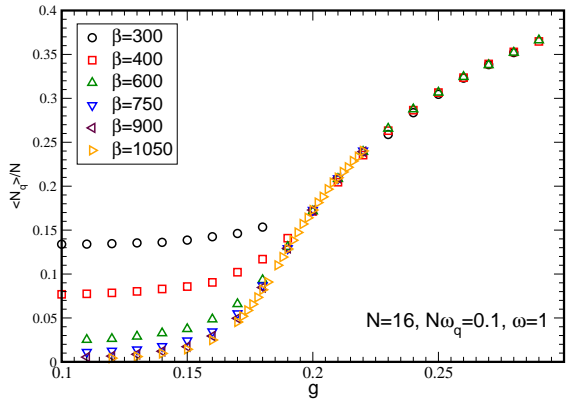}
\caption{(Color online) The order parameter, $\langle N_q\rangle/N$, 
obtained with QMC, versus the coupling, $g$. We see that for small $g$, $\beta$ 
must be very large for the system to probe the ground state. The error bars are 
	smaller than the symbol size. 
\label{dickeNqvsGbeta}}
\end{figure}

It is interesting to examine the accuracy of mean field calculations
in various situations. We therefore performed QMC simulations using
the stochastic Green function (SGF)
method\cite{rousseau08,rousseau08b}, both to verify our QMC
simulations and obtain the numerically exact ground state phase
diagram. To map out the phase diagram, we choose $\omega=1$ to set the
energy scale, and we calculate the order parameter for many values of
$g$ at fixed $N\omega_q$.  Making several such cuts for different
$N\omega_q$ yields the phase diagram in the $(N\omega_q, g)$ plane. In
the mean field calculation, the order parameter was ${\bar b}=\langle
GS|b|GS\rangle$; in the QMC simulations, we take the order parameter
to be $\langle N_q\rangle$, the average number of excited qubits which
corresponds to $|{\bar b}|^2$ in the mean field case. Figure~\ref{dickeNqvsGbeta} 
shows such a cut in $g$ for a $16$-particle
system at $N\omega_q=0.1$ and several values of the inverse
temperature, $\beta$. It is seen that the finite temperature effects
are very pronounced for small values of the coupling $g$, and that to
detect the transition properly, the temperature must be very low, and
gets lower with increasing $N$.

\begin{figure}[!h]
\includegraphics[width=1 \columnwidth]{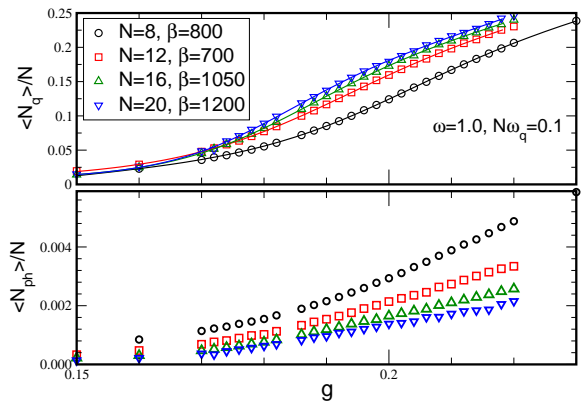}
\caption{(Color online) Top: The order parameter, $\langle
  N_q\rangle/N$, versus $g$ for several system sizes. The $\beta$
  values were chose such that the system is in the ground state. The
  solid curves going through the points are given by Pad\'e
  approximants (see text). Bottom: The average photon density versus
  $g$.
\label{dickeNqNphvsgN}}
\end{figure}

Next, we study the finite size effects on the behavior of the order
parameter.  This is presented in Fig.~\ref{dickeNqNphvsgN} where the
top panel shows the dependence of $\langle N_q\rangle/N$ on $N$
and $g$. The values of $\beta$ are chosen large enough so that the
system is in its ground state. It is seen that there is a change in
curvature as $g$ increases, and that the point of maximum slope shifts
to smaller values of $g$ as $N$ increases. On the other hand, the
lower panel shows that the average photon density does not exhibit any
sudden changes; it increases mildly with $g$. This was remarked with
the mean field results in Ref.~\onlinecite{garbe17}.

\begin{figure}[!h]
\includegraphics[width=1 \columnwidth]{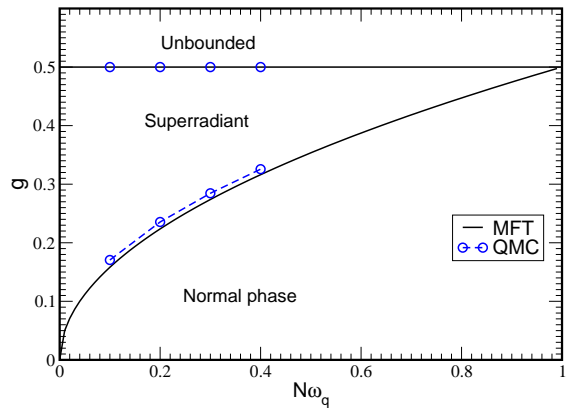}
\caption{(Color online) Phase diagram of the Dicke model given by mean field 
and SGF QMC simulations. The QMC results are extrapolated to the 
thermodynamic limit.
\label{dickephasediag}}
\end{figure}

In order to determine the transition point for each system size, we
fit Pad\'e approximants to the order parameter as a function of $g$
for each system size. The approximants we use are of the form,
\begin{equation}
 P(x) = \frac{a+bx+cx^2+dx^3}{1+ex+fx^2+hx^3};
\end{equation}
the maximum of the derivative yields the transition point for that
system size.  We then extrapolate the value of the critical $g$ to the
thermodynamic limit. To determine the boundary between the superradiant 
phase and the unstable region, we calculate $\langle N_{ph} \rangle/N$ as a 
function of $g$ for a fixed $N\omega_q$. We find that as $g\to 1/2$, the photon 
density increases very rapidly and becomes unmanageable at $g=1/2$ indicating 
the instability. This way we map the phase diagram shown in 
Fig.~\ref{dickephasediag} and which agrees very well with the MF solution. 
This excellent agreement can be understood as being due to the fact that 
the photon mode is coupled to all particles and introduces an effective long 
range interaction.

\section{The Jaynes-Cummings-Hubbard Model}

We now consider the JCH model given by Eq.~(\ref{rabihub}) but 
ignoring the CR terms,
\begin{eqnarray}
\nonumber
H_{JC}&=& -J \sum_{i=1}^N \left ( {\hat a}^\dagger_i {\hat 
a}^{\phantom\dagger}_{i+1} + h.c. \right ) + 
\sum_{i=1}^N
\left (\omega {\hat a}^\dagger_i {\hat a}^{\phantom\dagger}_i + 
\omega_q \sigma^+_i \sigma^-_i  \right )\\
&&+ g\sum_{i=1}^N \left ( \sigma^+_i {\hat 
a}^2_i + \sigma^-_i {\hat a}^{\dagger 2}_i\right).
\label{jchub}
\end{eqnarray}
In what follows, we set the energy scale by fixing $g=1$. 

\subsection{Perturbation}

We start in
the $J\ll 1$ limit where perturbation in $J$ can be expected to give
accurate results.

In the $J=0$ limit, the model can be solved exactly since the
eigenstates are dressed excitons labeled by the exciton number, $n$,
and upper or lower branch, $\pm$, which can be written as a
superposition of a Fock state with $n$ photons plus atomic ground
state, $|n,g\rangle$, and a Fock state with $n-2$ photons plus atomic
excited state, $\ket{n-2, e}$,
\begin{eqnarray}
 |n,+\rangle =\sin \theta_n |n,g\rangle +\cos \theta_n |n-2,e\rangle
 \nonumber \\
 |n,-\rangle =\cos \theta_n |n,g\rangle -\sin \theta_n
 |n-2,e\rangle
 \label{nbranch}
\end{eqnarray}
with the angle $\tan\theta_n = 2g\sqrt{n(n-1)}/(\Delta+R_n)$, 
$R_n(\Delta)=\sqrt{4g^2n(n-1)+\Delta^2}$, and the detuning parameter $\Delta 
= \omega_q-2\omega$.

\begin{figure}[!h]
\includegraphics[width=1 \columnwidth]{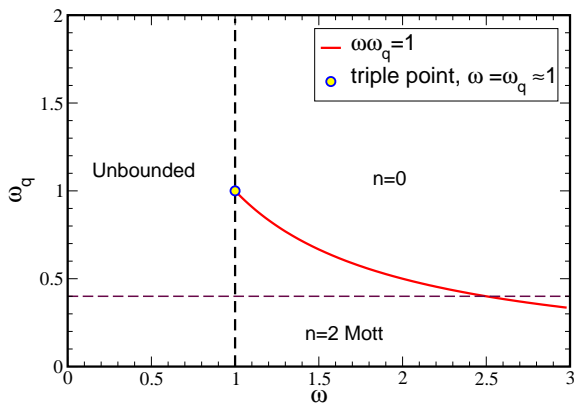}
\caption{(Color online) Phase diagram of the two-photon JC Hubbard model in 
the $J=0$ limit. The horizontal dashed line at $\omega_q=0.4$ will be discussed 
below.
\label{phasediagram_2pJCHJo0}}
\end{figure}

The corresponding eigenvalues are
\begin{equation}
 E_{n\pm}=n \omega+\Delta/2\pm R_n(\Delta)/2. 
 \end{equation}
The zero-exciton state $\ket{0 -}$ is a special case with
$E_{0-}=0$. The energy difference $R_n(\Delta)$ defines the Rabi splitting of
the upper and the lower branches. By comparing the energy of
neighboring exciton number $n=2$ , we obtain the border $\omega
\omega_q<1$ between vacuum and the $n=2$ Mott state.  Leaving the
$n=2$ Mott state or the vacuum at $\omega_q>1$ by increasing exciton
number, renders the Hamiltonian unbounded from below and makes the
system unstable at $\omega=1$. The $J=0$ phase diagram is shown in
Fig.~\ref{phasediagram_2pJCHJo0}.

At small nonzero tunneling, $J \ll 1$, we ignore the upper exciton
branch and perform a perturbation calculation to second order in $J$
to obtain the phase diagram of the JCH model. The ground state energy
of the Mott phase, $E_M$, that of the state with one additional
exciton, $E_d$, and that of the state with two additional excitons
$E_{2d}$ are calculated. We find that the energy at the line
determined by $E_M =E_d$ is smaller than $E_{2d}$ at the same $\omega$. 
This tells us that the upper Mott boundary is determined
by doping with a single exciton (adding one photon). The boundary is
given by,
\begin{equation}
\begin{split}
	&2\omega+(R_2-R_3) -4J t^{(3)2}_{--} +8J^2 \Biggl[ t_{--}^{(2)2}  \\
	&\left(-\sum_{\sigma=\pm}  \frac{2 t_{\sigma-}^{(3)2}}{\Delta-2R_2-\sigma R_3} 
+\frac{t_{--}^{(3)2} }{\Delta-2R_2+R_3}\right)\\
	&+\sum_{\sigma=\pm}\frac{t_{\sigma-}^{(4)2}}{\Delta-R_2-R_3-\sigma R_4}\\
	&+\frac{t_{+-}^{(3)2} t_{-+}^{(3)2}}{-2R_2-2R_3} -\frac{t_{+-}^{(3)2} 
t_{--}^{(3)2}}{R_3}-\frac{t_{--}^{(3)2}t_{-+}^{(3)2}}{-2R_2}\Biggr]=0, 
\label{upperbdry}
\end{split}
\end{equation}
where the matrix elements
\begin{equation}
	t^{(n)}_{\sigma\nu}=\bra{n,\sigma} {\hat a}^{\dagger}\ket{n-1,\nu},
\end{equation}
and $\sigma,\nu=\pm$, can be expressed in terms of the angle
$\theta_n$ and exciton number $n$. See Appendix \ref{perturbapp} for
details.

To determine the lower Mott boundary, we remove one exciton (single
holon doping) and calculate $E_{h}$, and also remove two excitons
(double holon doping) and calculate $E_{2h}$. The energy $E_h$
is always higher than $E_{2h}$  at $\omega$ determined by $E_M=E_{2h}$ indicating that the lower
boundary of the Mott phase is given by $E_M=E_{2h}$,
\begin{equation} 
\begin{split}
4\omega+\Delta-R_2 +16 J^2 t_{--}^{(2)} \biggl[ \frac{ 
t_{--}^{(3)2}}{\Delta+R_3-2R_2}\\
+\frac{t_{+-}^{(3)2}}{\Delta-R_3-2R_2} -\frac{1}{\Delta-R_2}\biggr]=0.
\end{split}
\label{lowerbdry}
\end{equation}
It is very interesting that, whereas single exciton doping determines
the upper boundary, two excitons should be removed to determine the
lower Mott boundary. This is confirmed by numerical calculations and
has other consequences which will be discussed below. The natures of
the phases above and below the Mott lobe will be discussed below.

The boundary between the vacuum and a state with at least two excitons
is determined by the equation $E_{2e}=E_{0-}=0$, where $E_{2e}$ is the
energy of the 2-exciton state:
\begin{eqnarray}
4w+\Delta-R_2+\frac{16J^2 t_{--}^{(2)2}}{\Delta-R_2}=0.
\label{vacbdry}
\end{eqnarray}
The resulting phase diagram, with the choice $\omega_q=0.4$, is shown in
Fig.~\ref{phasediagJCHperturb}.  Details of our perturbation
calculation are shown in Appendix \ref{perturbapp}. In addition, 
Fig.~\ref{phasediagJCHperturb} shows the mean field boundary (see Appendix \ref{mftjch} for details), 
Eq.~(\ref{mftjchunstab}), between stable thermodynamic phases and the 
unstable region where the Hamiltonian is unbounded from below. For comparison, 
we also show in Fig.~\ref{phasediagJCHperturb} the phase boundaries obtained 
with DMRG. As expected, we see very good agreement between perturbation and 
DMRG for small values $J$.

\begin{figure}[!h]
\includegraphics[width=1 \columnwidth]{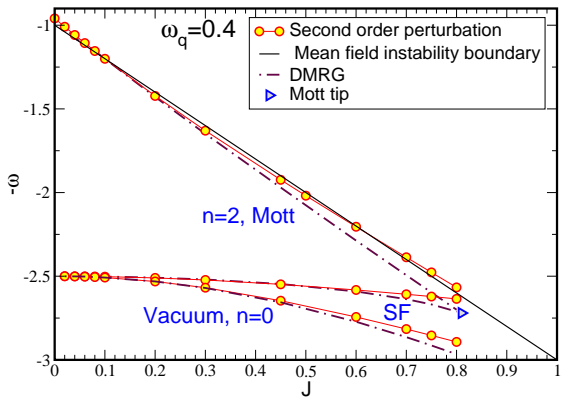}
\caption{(Color online) The phase diagram of the two-photon JCH
  model. The phase boundaries from second order perturbation are given
  by the circles. The mean field instability boundary is shown by the
  solid line. For comparison, we include the phase boundaries from
  DMRG calculations (see below). We do not show all the numerical
  results here to keep the figure uncluttered (see
  Fig.~\ref{phasediagJCHnum}).
\label{phasediagJCHperturb}}
\end{figure}

\subsection{Numerical results}

To obtain the phase diagram with exact numerical methods, we employ
the SGF\cite{rousseau08,rousseau08b} quantum Monte Carlo method and
DMRG\cite{white92,white93} using the ALPS\cite{alps}
library. The SGF and DMRG methods offer complementary advantages and
allow us to map out the phase diagram more precisely. Some typical
running times for the SGF QMC simulations are: $216$ hours for $N=24$
at $\beta=768$ and $192$ hours for $N=20$ at $\beta=640$.  These long
running times and large values of $\beta$ are necessary to ensure
convergence to the ground state. As for DMRG, we typically took the
maximum number of photons/site, $N_{max}=5$, $500$ states, and $500$
sweeps. These values are typical but the details depend on the
couplings and the phase of the system. In all cases, we verified that
increasing these values did not change the results.

\begin{figure}[!h]
\includegraphics[width=1 \columnwidth]{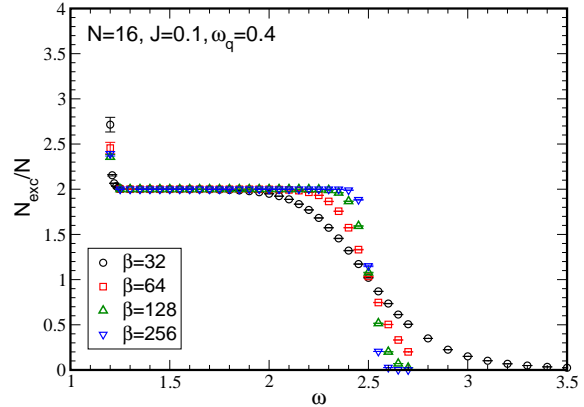}
\caption{(Color online) The dependence of the exciton density on the
  photon frequency, $\omega$. The horizontal line indicates the
  existence of an incompressible Mott insulating phase surrounded by
  two compressible phases.
\label{sgfcut1}}
\end{figure}

As in the previous section, here we set the energy scale by fixing $g=1$ and 
take $\omega_q=0.4$. We start with grand canonical QMC simulations using the 
SGF algorithm. Figure~\ref{sgfcut1} shows the density of excitons as a function 
of the photon frequency, $\omega$, for several values of the inverse 
temperature $\beta$. This allows us to determine how large $\beta$ needs to be 
to probe the ground state. Note that in the grand canonical ensemble, $\omega$ 
is minus the photon chemical potential. We see a clear 
incompressible Mott insulator (MI) plateau surrounded by compressible regions. 
We note that the MI becomes fully formed only for very large $\beta$. This is 
due to the small value of $J$ we chose for the figure; larger values of $J$ do 
not require such high values of $\beta$, but, of course, at larger $J$, the MI 
is not as wide (see Fig.~\ref{phasediagJCHperturb}).

\begin{figure}[!h]
\includegraphics[width=1 \columnwidth]{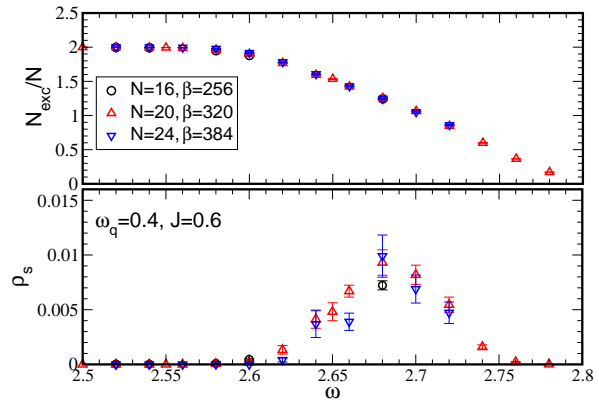}
\caption{(Color online) Top: The exciton density as a function of
  $\omega$.  Bottom: The corresponding superfluid density, $\rho_s$,
  as a function of $\omega$. This shows that as the MI is doped with
  holes, a superfluid phase appears.
\label{sgfcut2}}
\end{figure}

\begin{figure}[!h]
\includegraphics[width=1 \columnwidth]{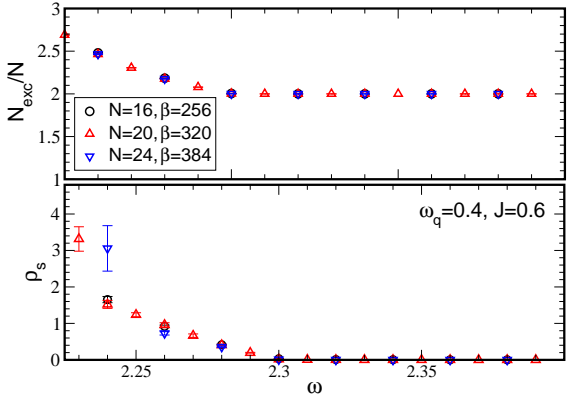}
\caption{(Color online) Same as Fig.~\ref{sgfcut2} but doping the MI with 
particles. A superfluid phase appears here too.
\label{sgfcut3}}
\end{figure}

\begin{figure}[!h]
\includegraphics[width=1 \columnwidth]{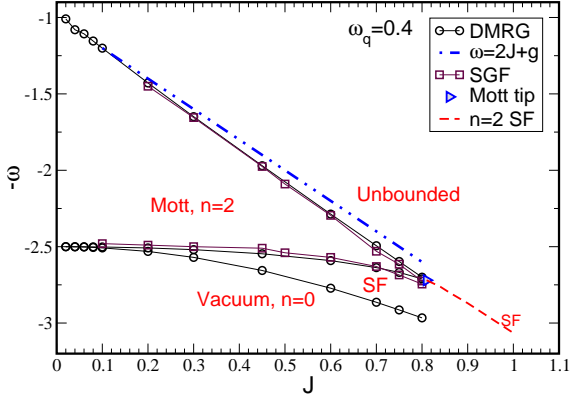}
\caption{(Color online) The phase diagram of the JCH model from SGF
  QMC and DMRG calculations. Dash-double dot line ($\omega=2J+g$) is
  the MF boundary between the unstable and stable regions. The (blue)
  right triangle indicates the position of the tip of the Mott
  lobe. The dashed (red) line indicates the SF phase with a density of
  $2$ excitons/site. This is the same as Fig.~\ref{phasediagJCHperturb}
  but not showing the perturbation results for clarity.
\label{phasediagJCHnum}}
\end{figure}

To delineate the nature of the compressible regions surrounding the MI, we show 
in Figs.~\ref{sgfcut2} and \ref{sgfcut3} the exciton density as a function of 
$\omega$ in the top panel, and the superfluid density, $\rho_s$, in the bottom 
panel. We see in both figures that, as soon as the system leaves the MI phase, 
it becomes superfluid. Hole doping, Fig.~\ref{sgfcut2}, results in a 
very dilute SF before the system eventually becomes empty as $\omega\approx 
2.8$. Particle doping, Fig.~\ref{sgfcut3}, leads to a SF with $\rho_s$ 
approaching $2$ before the system starts to become unstable at $\omega \approx 
2.23$. By making several cuts of this type, we map out the phase diagram shown 
in Fig.~\ref{phasediagJCHnum} obtained numerically with very good agreement 
between SGF and DMRG. The figure shows a single Mott lobe with $2$ 
excitons/site; below this lobe there is a rather narrow region of SF before the 
system becomes empty. Above the Mott lobe, there is an even narrower strip of SF 
before the system becomes unstable. It is hard to pinpoint precisely with SGF 
and DMRG the boundary where the system becomes unstable because the number of 
photons/site increases very rapidly near the instability. Neither DMRG nor QMC 
performs well under such conditions and, for that reason, we show in the figure 
the stable/unstable boundary given by MF calculations. Unlike the case of the 
1-photon JCM\cite{rossini07}, there are no higher Mott lobes in this 2-photon 
case due to the instability triggered by the Hamiltonian becoming unbounded 
from below. The Mott lobe terminates in a cusp because the critical point at 
the tip is in the BKT universality class.

\begin{figure}[!h]
\includegraphics[width=1\columnwidth]{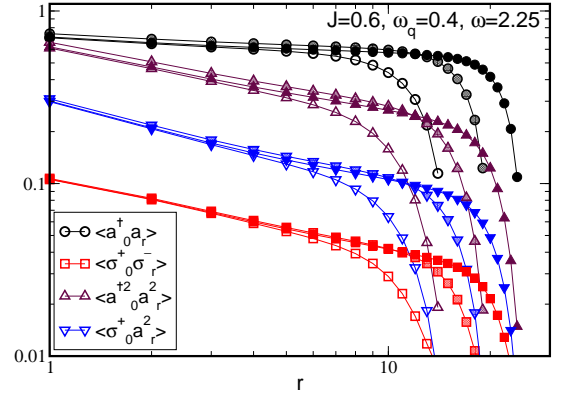}
\caption{(Color online) Photon-photon, photon-qubit and qubit-qubit Green 
functions from DMRG in the SF region above the MI. The open, shaded and full 
symbols are for $N=30,40$ and $50$ respectively.
\label{gfunctsaboveMI}}
\end{figure}

\begin{figure}[!h]
\includegraphics[width=1\columnwidth]{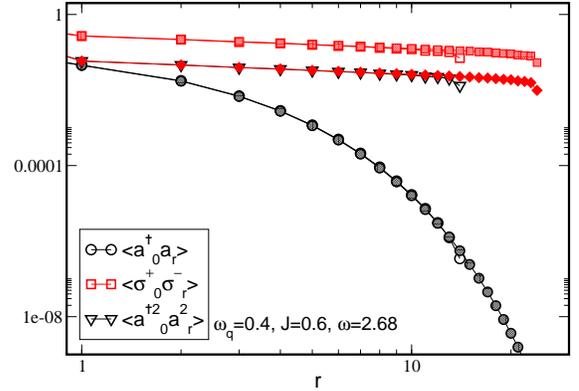}
\caption{(Color online) Same as Fig.~\ref{gfunctsaboveMI} but below the 
Mott lobe. The open, shaded symbols are for $N=30,50$ respectively.
\label{gfunctsbelowMI}}
\end{figure}

The question arises as to the nature of the SF phase. In the 1-photon
system\cite{rossini07}, the photon Green function, $\langle {\hat
  a}^{\dagger}_0 {\hat a}^{\phantom\dagger}_{r}\rangle$ decays as a
power indicating quasi-long range order and a photon SF. At the same
time, the photon-qubit Green function, $\langle \sigma^+_0 {\hat
  a}^{\phantom\dagger}_{r}\rangle$, also decays as a power law. We see
the same behavior in Fig.~\ref{gfunctsaboveMI} where several two-point
functions are shown on log-log scale. We note, in particular, that
$\langle {\hat a}^{\dagger}_0 {\hat a}^{\phantom\dagger}_{r}\rangle$
decays as a power law, and that, while both $\langle {\hat
  a}_0^{\dagger 2}{\hat a}_r^2\rangle$ and $\langle \sigma^+_0 {\hat
  a}^2_r\rangle$ decay as powers, they are not equal. The leading effect in 
this phase is then the coherent movement of individual photons.

One expects to encounter similar behavior below the Mott lobe. However, 
this expectation is not borne out by the numerical results. Figure~\ref{gfunctsbelowMI} 
shows the same correlation functions as 
Fig.~\ref{gfunctsaboveMI} where we see clearly that, although the system is in 
the SF phase, the quantity $\langle {\hat a}_0^\dagger {\hat a}_r\rangle $ 
decays exponentially while it exhibited power law behavior above the MI. The 
correlation functions $\langle \sigma^+_0 \sigma^-_r\rangle $ and $\langle 
{\hat a}^{\dagger 2}_0 {\hat a}^2_r\rangle$ both decay as 
powers showing the SF nature of 
this phase. In addition, we find that, unlike the case above the MI (see 
Fig.~\ref{gfunctsaboveMI}), the correlation function $\langle \sigma^+_0 {\hat 
a}^2_r \rangle$ is {\it identical} to $\langle {\hat a}^{\dagger 2}_0 {\hat 
a}^2_r\rangle $, and so we do not show it in Fig.\ref{gfunctsbelowMI} to 
keep the figure clear.

This behavior appears to show that above the MI, we have a photon SF phase 
(power law decay for $\langle {\hat a}^\dagger_0 {\hat 
a}^{\phantom\dagger}_r\rangle$) whereas below the MI we have a photon {\it 
pair} SF phase (exponential decay of $\langle {\hat a}^\dagger_0 {\hat 
a}^{\phantom\dagger}_r\rangle$ and power law decay of $\langle {\hat 
a}^{\dagger 2}_0 {\hat a}^2_r\rangle$): Below the Mott insulator the photons 
appear to form bound pairs which become superfluid.

\section{The Rabi-Hubbard model}

\begin{figure}[!h]
\includegraphics[width=1\columnwidth]{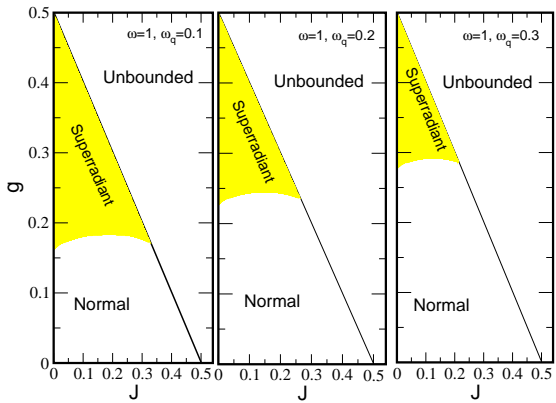}
\caption{(Color online) The mean field phase diagram of the RH model exhibiting 
normal and superradiant phases and an unbounded region. This behavior is 
similar to that of the Dicke model.
\label{rhmmftphase}}
\end{figure}

In Appendix \ref{mftapp} we show the mean field calculation for the 
Rabi-Hubbard model predicting a phase transition between a disordered 
(normal) phase and a superradiant phase similar to the Dicke 
model\cite{garbe17}. The boundary between the two phases is given by (see 
Eq.~(\ref{egmft1}))
\begin{eqnarray}
E_G({\bar b}\neq0, g, J, \omega, \omega_q)=E_G({\bar b}=0, g, J, \omega, 
\omega_q),
\label{mftRHphase}
\end{eqnarray}
where $E_G$ is the ground state energy given by (see Eq.~(\ref{egmft1})
\begin{eqnarray}
E_G&=&\frac{1}{2}\sum_k\biggl[\sqrt{(-2J\cos(k)+\omega)^2-16g^2{\bar
      b}^2+16g^2{\bar b}^4} \nonumber \\ &+&\omega_q{\bar
    b}^2-\frac{\omega}{2}\biggr]. \nonumber \\
\label{mftEg}
\end{eqnarray}
In addition, the stability condition requires (see Eq.~(\ref{stabcondtion}),
\begin{equation}
\omega -2J\geq 2g.
\end{equation}
The two phases and the unstable region are shown in
Fig.~\ref{rhmmftphase}. This phase diagram is the analog of that of the
Dicke model, Fig.~\ref{dickephasediag}. However, whereas in the case of
the Dicke model, we found excellent agreement between the MF and QMC
phase diagrams (Fig.~\ref{rhmmftphase}), this is not the case for the
RH model.

\begin{figure}[!h]
\includegraphics[width=1\columnwidth]{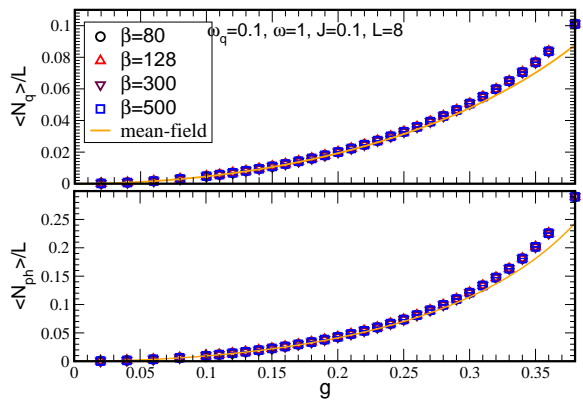}
\caption{(Color online) Excited qubit density, $\langle N_q \rangle /N$, (upper panel)
  and photon density (lower panel) as functions of $g$. Various
  $\beta$ values are used to make sure the system is in the ground
  state.
\label{rhmNqvsgbeta}} 
\end{figure}
\begin{figure}[!h]
\includegraphics[width=1\columnwidth]{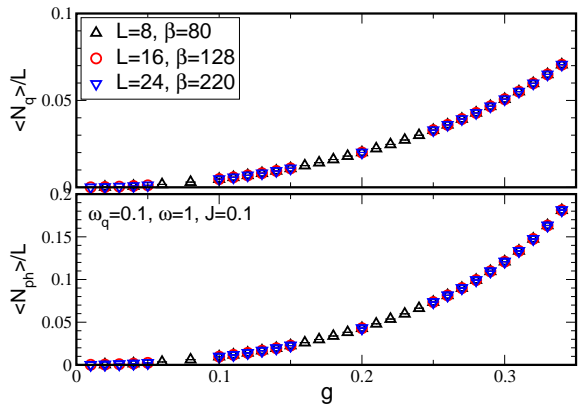}
	\caption{(Color online) The same as fig. \ref{rhmNqvsgbeta}
          but for different system sizes showing that finite size
          effects are negligible.
\label{rhmNqvsgL}}
\end{figure}

We show in Fig.~\ref{rhmNqvsgbeta} the average number of excited
qubits, $\langle N_q\rangle/N $, (top) and the average photon density,
$\langle N_{ph}\rangle /N$, (bottom) as functions of the coupling $g$
for several values of $\beta$. We observe a smooth continuous increase
of both quantities as $g$ increases and no finite temperature
effects. The values are in good agreement with the mean-field approach
described in Appendix~\ref{mftjch}, which, for the parameters of
Fig.~\ref{rhmNqvsgbeta}, predicts a direct transition from a normal
phase to the unbounded region. On the other hand, since the same
mean-field approach applied to the JCH Hamiltonian is not able to
capture the photon pair SF phase, it is possible that it also misses
an SF phase in the RH model for parameters different from shown in the
figure.  In Fig.~\ref{rhmNqvsgL}, we show that for sufficiently large
$\beta$ there is no finite size effects. Comparing the top panels of
Figs.~\ref{rhmNqvsgbeta} and \ref{rhmNqvsgL} with
Fig.~\ref{dickeNqvsGbeta} emphasizes the difference in behavior
between the two models and suggests the absence of a phase transition
for the RH model, at least for the present parameter values. For the
entire range of values of $g$ in Figs.\ref{rhmNqvsgbeta} and
\ref{rhmNqvsgL}, all Green functions, such as $\langle {\hat
  a}^\dagger_0 {\hat a}_r\rangle$ and $\langle \sigma^+_0 {\hat a}^2_r
\rangle $, decay exponentially indicating a normal rather than a
superradiant phase.

In short, at least for the parameter values we have considered, the
Rabi-Hubbard model seems not to exhibit a phase transition, even
though the underlying discrete $Z_4$ symmetry allows it. In addition,
we have two different mean-field approaches leading to rather
different phase diagrams, which seems to indicate that the physics of
the Rabi-Hubbard model might be more involved than the one of the
Dicke and the JC models. A thorough study is beyond the scope of the
present paper and will be presented elsewhere.

\section{Conclusions}

In this paper we studied the ground state properties of the Dicke,
Jaynes-Cummings Hubbard, and Rabi-Hubbard models. We used mean field,
perturbation, QMC, and DMRG calculations to elucidate the phase
diagrams and the transitions between the various phases. We found that
for the Dicke model, exact QMC results agree very well with the mean
field phase diagram\cite{garbe17} but that very large $\beta$ was
required to probe the ground state properties. For the JCH model, we
found that at small hopping, $J$, the system exhibits a single
incompressible Mott insulator with two excitons/site. Doping above the
MI, the system exhibits a photon SF phase before becoming unstable due
to the Hamiltonian becoming unbounded from below.  To dope the system
above the MI, we add one photon at a time. This is in big contrast to
the situation below the MI where we found a surprising photon {\it
  pair} SF: The photons pair up in bound states. Below the MI, one
needs to remove two photons at a time to dope. For the JCH model in
the SF phase, very large values of $\beta$ are needed. Interestingly,
we have not found a phase transition between a normal and a
superradiant phase in the case of the RH model; only a normal phase with 
exponential decay of the Green functions. It would be
interesting to study how the three-photon model would differ from what
we found here. In particular,would the JCH model still exhibit a MI
phase, and would doping below it yield a three-photon bound state
superfluid.

%%%%%%%%%%%%%%%%%%%%%%%%%%%%%%%%%%%%%%
\begin{acknowledgments}
S.C and W.G. was supported by the NSFC under Grant No.~11775021 and
No.~11734002.
\end{acknowledgments}

\appendix
%%%%%%%%%%%%%%%%%%%%%%%%%%%%%%%%%%%%%%%%%%%%%%%%%%%%%%
\section{Mean field two-photon RH}
\label{mftapp}
In this appendix we outline the mean field calculation of the phase
diagram of the two-photon Rabi-Hubbard model which proceeds along the
same lines as that for the Dicke model in \onlinecite{garbe17}. The RH
Hamiltonian is:
\begin{eqnarray}
\nonumber
 H_{RH}&=& -J\sum_i \left ( {\hat a}^\dagger_i {\hat a}^{\phantom\dagger}_{i+1} 
+ h.c. \right )+ \sum_i \left (\omega_q 
\sigma^+_i\sigma^-_i + \omega {\hat a}^\dagger_i 
{\hat a}^{\phantom\dagger}_{i}\right )\\
&&+g \sum_i \left ( \sigma^+_i + \sigma^-_i \right ) \left ( {\hat 
a}^{\dagger\, 2}_i + {\hat a}^{2}_i \right ).
\label{rhm}
\end{eqnarray}
We apply the Holstein-Primakoff transformation:
\begin{eqnarray}
 \sigma^+_i &=&{\hat b}^\dagger_i \sqrt{1-{\hat b}^\dagger_i 
{\hat b}^{\phantom\dagger}_i}\\
  \sigma^-_i &=& \sqrt{1-{\hat b}^\dagger_i 
{\hat b}^{\phantom\dagger}_i}\,{\hat b}^{\phantom\dagger}_i,
\end{eqnarray}
with $[{\hat b}^{\phantom\dagger}_i,{\hat b}^\dagger_j]=\delta_{ij}$, 
use the mean field approximation,
\begin{eqnarray}
{\bar b}^*&=& \langle GS|{\hat b}^\dagger_i|GS \rangle,\\
{\bar b}&=& \langle GS|{\hat b}^{\phantom\dagger}_i|GS \rangle,\\
{\bar b}&=&{\bar b}^*,
\end{eqnarray}
and ignore fluctuations in the ${\hat b}$ field. This 
leads to the quadratic mean field Hamiltonian,
\begin{eqnarray}
\nonumber
 H_{MF}&=& -J\sum_i \left ( {\hat a}^\dagger_i {\hat a}^{\phantom\dagger}_{i+1} 
+ 
{\hat a}^\dagger_{i+1} {\hat a}^{\phantom\dagger}_{i} \right )+ \omega \sum_i 
{\hat a}^\dagger_i {\hat a}^{\phantom\dagger}_{i} \\
\nonumber
&&+2g{\bar b}\sqrt{1-{\bar b}^2} \sum_i \left ( {\hat a}^{\dagger\, 2}_i + 
{\hat a}^{2}_i \right )\\
&&+N_s\omega_q{\bar b}^2.
\label{rhmmf}
\end{eqnarray}
Applying a Fourier transform,
\begin{equation}
 {\hat a}_i= \frac{1}{\sqrt{N}} \sum_i {\rm e}^{-ikr_i} {\hat a}_k,
\end{equation}
leads to,
\begin{eqnarray}
\nonumber
 H_{MF}&=&\sum_k \bigg [ \left (-2J\cos(k) + \omega\right ){\hat 
a}^\dagger_k {\hat a}^{\phantom\dagger}_k\\
\nonumber
&&+2g{\bar b}\sqrt{1-{\bar b}^2}\left ({\hat a}^\dagger_k {\hat a}^\dagger_{-k}+ 
{\hat a}^{\phantom\dagger}_k {\hat a}^{\phantom\dagger}_{-k} \right ) \bigg ]\\
&&+N_s\omega_q{\bar b}^2.
\label{rhmmomspace}
\end{eqnarray}
Now, we use the Bogoliubov transformation:
\begin{eqnarray}
{\hat a}^\dagger_k &=& {\rm cosh}(X_k) {\hat \alpha}^\dagger_k - {\rm
  sinh}(X_k){\hat \alpha}^{\phantom\dagger}_{-k}\\
{\hat a}^{\phantom\dagger}_k &=& {\rm cosh}(X_k) {\hat
  \alpha}^{\phantom\dagger}_k- {\rm sinh}(X_k){\hat
  \alpha}^{\dagger}_{-k}.
\label{bogo}
\end{eqnarray}
To simplify the expressions, we write: $c_k\equiv {\rm cos}(k)$,
$ch_k\equiv {\rm cosh}(X_k)$, and $sh_k\equiv{\rm sinh}(X_k)$. This
yields,
\begin{eqnarray}
\nonumber
 H_{MF}&=& \sum_k \bigg [ \left (-2Jc_k + \omega\right ) \left ( ch_k^2 + 
sh_k^2\right )\\
\nonumber
&&-8g{\bar b}\sqrt{1-{\bar b}^2}\,ch_k sh_k \bigg 
]{\hat \alpha}^\dagger_k{\hat \alpha}^{\phantom\dagger}_k\\
\nonumber
&+&\sum_k \bigg [ \bigg (2g{\bar b}\sqrt{1-{\bar b}^2}\,(ch_k^2+sh_k^2) \\
\nonumber
&&-(-2Jc_k+\omega)ch_ksh_k \bigg ){\hat 
\alpha}^\dagger_k{\hat \alpha}^\dagger_{-k} + {\rm h.c.} \bigg] \\
\nonumber
&+&\sum_k \bigg [(-2Jc_k + \omega )sh_k^2 \\
&&- 4g{\bar b}\sqrt{1-{\bar b}^2}\,ch_k 
sh_k \bigg ] + N_s\omega_q{\bar b}^2.
\label{diagmfham}
\end{eqnarray}
The first term in Eq.~(\ref{diagmfham}) gives the quasi-particle
excitations above the ground state; the second term contains the
nondiagonal terms which we want to eliminate and the third term gives
the ground state energy. When the coefficient of the nondiagonal term
is made to vanish, we obtain the condition:
\begin{equation}
2g{\bar b}\sqrt{1-{\bar b}^2}\,(ch_k^2+sh_k^2) =
(-2Jc_k+\omega)ch_ksh_k,
\end{equation}
and using
\begin{eqnarray}
\nonumber {\rm sinh}(2x)&=& 2{\rm sinh}(x){\rm cosh}(x),\\ {\rm
  cosh}(2x)&=& {\rm sinh}^2(x)+{\rm cosh}^2(x),
\end{eqnarray}
we get:
\begin{equation}
{\rm tanh}(2X_k) = \frac{4g{\bar b}\sqrt{1-{\bar b}^2}}{\omega-2J{\rm
    cos}(k)}.
\label{tanhxk}
\end{equation}
Writing
\begin{equation}
 A=2g{\bar b}\sqrt{1-{\bar b}^2},\,\,\,\,\,\,B_k = \omega-2J{\rm
   cos}(k),
\label{abk}
 \end{equation}
gives:
\begin{equation}
 {\rm tanh}(2X_k) = \frac{2A}{B_k},
 \label{tanhcond}
\end{equation}
and,
\begin{equation}
 {\rm e}^{X_k} = \left [ \frac{B_k+2A}{B_k-2A}\right ]^{1/4}.
 \label{expcond}
\end{equation}
The acceptable values satisfy the condition:
\begin{equation}
 -1 \leq \frac{2A}{B_k} \leq 1.
\end{equation}
Then, the energy that must be minimized with respect to ${\bar b}$ is
\begin{eqnarray}
\nonumber
 E_G&=& \sum_k \bigg [(-2Jc_k + \omega )sh_k^2 - 
4g{\bar b}\sqrt{1-{\bar b}^2}\,ch_k 
sh_k \bigg ]\\
&&+ N_s\omega_q{\bar b}^2.
\label{Egrdstate}
\end{eqnarray}
Note that, putting $J=0$, renders all the hyperbolic functions
independent of $k$. Then Eq.~(\ref{Egrdstate}) agrees with the MF
equation in Ref.~\onlinecite{garbe17} if we take $N=1$ and ${\bar
  b}={\bar b}^*$.

Equation~(\ref{Egrdstate}) can be further simplified to the form,
\begin{eqnarray}
E_G&=&\frac{1}{2}\sum_k\biggl[\sqrt{(-2J\cos(k)+\omega)^2-16g^2{\bar
      b}^2+16g^2{\bar b}^4} \nonumber \\ &+&\omega_q{\bar
    b}^2-\frac{\omega}{2}\biggr]. 
\label{egmft1}
\end{eqnarray}
The requirement that the argument of the square root be positive leads
to the important condition:
\begin{equation}
\omega -2J\geq 2g.
\label{stabcondtion}
\end{equation}
This is the stability condition for the system: When it is violated,
the Hamiltonian is unbounded from below.

The transition to the superradiant phase is indicated by a nonzero
order parameter of the order parameter, ${\bar b}\neq0$. The boundary
between the normal and superradiant phases is, therefore, given by:
\begin{eqnarray}
E_G({\bar b}\neq0, g, J, \omega, \omega_q)=E_G({\bar b}=0, g, J, \omega, 
\omega_q).\quad
\label{egmft2}
\end{eqnarray}
This relation cannot be solved analytically, we solve it numerically to 
get the relation between $g, J, \omega, \omega_q$ which gives the phase 
boundary.

%%%%%%%%%%%%%%%%%%%%%%%%%%%%%%%%%%%%%%%%%%%%%%%%%%%%%%

\section{Simple mean field for JCH}
\label{mftjch}
The mean field calculation in Appendix \ref{mftapp} applies to the
Rabi-Hubbard model and, by ignoring the counter-rotating terms, also
to the JCH model.  However, we can also perform the mean field
calculation on the hopping term of the photon field rather than on the
qubit.  We assume $\langle a_i \rangle = \psi_i$, which reduces the
Hamiltonian to a single-site form:
\begin{equation}
\begin{aligned}
 H_i^{MF}=&-J\left((\psi_{i+1}+\psi_{i-1})a^{\dagger}_i+\text{h.c.}\right)\\
 &+\omega_q\sigma_i^+\sigma_i^-+\omega a^{\dagger}_i a_i+
 g\left ( \sigma^+_i + \sigma^-_i \right ) \left ( {\hat 
a}^{\dagger\, 2}_i + {\hat a}^{2}_i \right ).
\end{aligned}
\end{equation}
Finding the ground state of this Hamiltonian and a self-consistent
solution for the $\psi_i$ allows us to obtain the phase diagram. In
particular, in the normal phase $\psi_i=0$, the ground state of
$H_i^{MF}$ still depends on the parameters $\omega_q$, $\omega$ and
$g$, leading, for instance, to the dependence of the qubit and photon
densities on the interaction strength observed in
Fig.\ref{rhmNqvsgbeta}.

A further approximation amounts to replacing all the photonic
operators by the mean field value, $\psi$, thus yielding the leading
order contribution which gives the instability line for the JCH
model. In that case, assuming that $\psi$ is real, the mean field
Hamiltonian becomes
\begin{eqnarray}
H_i^{SMF}&=&\omega_q\sigma_i^+\sigma_i^- 
+\omega\psi^2+g(\sigma_i^++\sigma_i^-)\psi^2-2J\psi^2. \nonumber \\
\end{eqnarray}
The ground state energy is found by minimizing
\begin{eqnarray} 
E_G=\frac{\omega_q}{2}+(\omega-2J)\psi^2-\frac{\sqrt{\omega_q^2+4g^2\psi^4}}{2},
\end{eqnarray}
with respect to $\psi$ which gives
\begin{eqnarray}
\omega-2J=\frac{2g^2\psi^2}{\sqrt{\omega_q^2+4g^2\psi^4}}.
\label{dEdpsi}
\end{eqnarray}
Near the unbounded region, $\psi \gg 1$ and we can thus ignore the 
$\omega_q$ term and find that for stability we must have,
\begin{equation}
\omega-2J \geq g.
\end{equation}
The boundary of the unstable region is therefore given by,
\begin{equation}
 \omega_c - 2J=g.
 \label{mftjchunstab}
\end{equation}
Note the similarity between this equation and the stability condition for the 
RHM, Eq.~(\ref{stabcondtion}).

We then define the photon density $n_{ph}=\psi^2$ and, using Eq.~(\ref{dEdpsi}) 
we obtain,
\begin{eqnarray}
n_{ph}=\frac{\omega_q(\omega-2J)}{2g}\frac{1}{\sqrt{(g-\omega+2J) 
(g+\omega-2J)}}.\quad
\end{eqnarray}
Now we consider $\omega$ near the stable-unstable boundary,
$\omega_c$, so that $\omega-2J\approx g$. We find that the number of photons 
diverges as the unstable region is approached:
\begin{equation}
n_{ph}= \frac{\omega_q}{2\sqrt{2g}}\frac{1}{\sqrt{2J+g-\omega}} \propto
(\delta \omega)^{1/2},
\label{nph}
\end{equation}
with $\delta \omega =\omega_c-\omega$. This divergence makes the result 
difficult to demonstrate with QMC and/or DMRG methods.

%%%%%%%%%%%%%%%%%%%%%%%%%%%%%%%%%%%%%%%%%%%%%%%%%%%%%%
\section{Perturbation calculation of the phase diagram of the JCH model}
\label{perturbapp}

In this appendix we outline the perturbation calculation of the phase
diagram of the two-photon JCH model which proceeds along the same
lines as that for the single-photon JCH model in
\onlinecite{schmidt09}.

The Hamiltonian of the two-photon JCH model is
\begin{eqnarray}
\nonumber
H_{JC}&=& -J \sum_{i=1}^N \left ( {\hat a}^\dagger_i {\hat 
a}^{\phantom\dagger}_{i+1} + h.c. \right ) + 
\sum_{i=1}^N
\left (\omega {\hat a}^\dagger_i {\hat a}^{\phantom\dagger}_i + 
\omega_q \sigma^+_i \sigma^-_i  \right )\\
&&+ g\sum_{i=1}^N \left ( \sigma^+_i {\hat 
a}^2_i + \sigma^-_i {\hat a}^{\dagger 2}_i\right).
\label{jchubapp}
\end{eqnarray}
which can be split into two parts,
\begin{eqnarray}
 H_{JC}=H_0+H_1,
\end{eqnarray}
with the perturbation, $H_1$, given by
\begin{equation}
H_1=-J\sum_i(a^\dagger_i a_{i+1}+ h.c.),
\end{equation}
and $H_0$ is the rest of $H$. 

It is convenient to introduce the matrix elements
\begin{equation}
t^{(n)}_{\sigma\nu}\equiv\bra{n,\sigma} a^{\dagger}\ket{n-1,\nu},
\end{equation}
with $\sigma, \nu=\pm$. 
Using Eq.~(\ref{nbranch}), we have
\begin{eqnarray}
%a|n,+>&=&t_{n++}|n-1,+>+t_{n+-}|n-1,-> \nonumber \\
%a|n,->&=&t_{n-+}|n-1,+>+t_{n--}|n-1,-> \nonumber \\
a^\dagger \ket{n-1,+}&=&t^{(n)}_{++}\ket{n,+} + t^{(n)}_{-+}\ket{n,-}, \nonumber \\
a^\dagger \ket{n-1,-} &=& t^{(n)}_{+-}\ket{n,+} + t^{(n)}_{--}\ket{n,-},
\end{eqnarray}
with  
\begin{equation}
t^{(n)}_{\sigma \nu}=\sqrt{n} \alpha^{\sigma}_n \alpha^{\nu}_{n-1}
+\sigma \nu \sqrt{n-2}\alpha_{n}^{-\sigma}\alpha_{n-1}^{-\nu},
\end{equation}
where
\begin{eqnarray}
&&\alpha_n^+=\sin \theta_n, \nonumber \\
&&\alpha_n^-=\cos \theta_n.
%&&b_n^+=\cos(\theta_n) \nonumber \\
%&&b_n^-=-\sin(\theta_n)
\end{eqnarray}

For the upper boundary of the $n=2$ Mott phase, we compare the ground
state energy of the $n=2$ Mott state with that of the state doped by one
exciton.  However, to determine the lower boundary of the $n=2$ Mott
phase, we need to compare the ground state energy with the state doped
by 2 holons (with two excitons removed from the system).  Similarly,
to locate the boundary of vacuum, we need to compare the energy of the
state doped by two excitons with vacuum.

We first calculate the energy of the Mott state to second order,
 \begin{eqnarray}
E_{M}=E_M^{(0)}+E_M^{(1)}+E_M^{(2)}, 
\end{eqnarray}
with
\begin{eqnarray}
E_M^{(0)}&=&\bra{\psi_M^{(0)}} H_0 \ket{\psi_M^{(0)}}=N(2 \omega 
+\frac{\Delta}{2}-\frac{R_2}{2}), \nonumber\\
E_M^{(1)}&=&\bra{\psi_M^{(0)}} H_1\ket{\psi_M^{(0)}}=0,\\
E_M^{(2)} &=& \bra{\psi_M^{(1)}} H_1\ket{\psi_M^{(0)}}
= 4J^2N\sum_{\sigma=\pm}\frac{t_{\sigma-}^{(3)2} t_{--}^{(2)2}}{\Delta-\sigma 
R_3-2R_2}, \nonumber
\end{eqnarray}
where the zero-th order wavefunction is
\begin{equation}
 \ket{\psi_{M}^{(0)}}=\prod^N_i \ket{2,-}_i,
\end{equation}
and the first order wavefunction is
\begin{eqnarray}
&&\ket{\psi_{M}^{(1)} }=\sum_k \frac{\ket{k}\bra{k}
    H_1\ket{\psi_{M}^{(0)}}} {E_{M}^{(0)}-E_k} \nonumber \\ &&=-J
  \sum_i \sum_{\sigma=\pm}\Biggl[\frac{t_{\sigma-}^{(3)}t_{--}^{(2)}
      \ket{1,-}_i \ket{3,\sigma}_{i+1} \prod^N_{l \neq i, l \neq i+1}
      \ket{2,-}_l}{\Delta-\sigma R_3-2R_2} \nonumber
    \\ &&+\frac{t_{\sigma-}^{(3)} t_{--}^{(2)} \ket{1,-}_{i+1}
      \ket{3,\sigma}_i \prod^N_{l\neq i, l\neq i+1}
      \ket{2,-}_l}{\Delta-\sigma R_3-2R_2}\Biggr].
\end{eqnarray}

Now we calculate the lowest energy of the state obtained by doping the $n=2$ 
Mott state, to second order,
\begin{eqnarray}
E_{d}=E_{d}^{(0)}+E_{d}^{(1)}+E_{d}^{(2)},
\end{eqnarray}
with
\begin{eqnarray}
E_{d}^{(0)} &=& \bra{\psi_d^{(0)}} H_0 \ket{\psi_d^{(0)}}\nonumber \\
&=&(N-1)(2\omega + \frac{\Delta}{2}-\frac{1}{2}R_2)+(3\omega + 
\frac{\Delta}{2}-\frac{R_3}{2}), \nonumber \\ 
E_{d}^{(1)} &=& \bra{\psi_{d}^{(0)}} H_1 \ket{ \psi_{d}^{(0)}}  = -2J 
t_{--}^{(3)2}, \nonumber \\ 
E_{d}^{(2)} &=& \bra{\psi_{d}^{(1)}} H_1 \ket{ \psi_{d}^{(0)}}  \nonumber \\ 
&=&4J^2\Biggl[\frac{t_{--}^{(3)2} 
t_{--}^{(2)2}}{\Delta+R_3-2R_2} +\sum_{\sigma=\pm}(\frac{(N-2) 
t_{\sigma-}^{(3)2} t_{--}^{(2)2}}{\Delta-2 R_2-\sigma R_3} \nonumber \\ 
&+& \frac{t_{\sigma-}^{(4)2} t_{--}^{(2)2}}{\Delta-R_2-R_3-\sigma R_4}
+\frac{t_{+-}^{(3)2} t_{-\sigma}^{(3)2}}{R_3-R_2-\sigma R_2}  \nonumber \\
&+& \frac{t_{\sigma(-\sigma)}^{(3)2} t_{--}^{(3)2}}{-R_3-\sigma R_3-R_2+\sigma 
R_2} )\Biggr],
\end{eqnarray}
where 
\begin{eqnarray}
\ket{\psi_{d}^{(0)}}=\frac{1}{\sqrt{N}} \sum_i  \ket{3,-}_i \prod^N_{l\neq 
i}\ket{2,-}_l, 
\end{eqnarray}
and
\begin{equation}
\ket{\psi_{d}^{(1)}}=\ket{\psi_d^{(1)}(1)}+\ket{\psi_d^{(1)}(2)}+\ket{\psi_d^{
(1)}(3)}, 
\end{equation}
 with
\begin{eqnarray}
\ket{\psi_{d}^{(1)}(1)} &=& \frac{1}{\sqrt{N}}\sum_{i, j, \sigma=\pm} 
\frac{-2J}{\Delta-(\sigma)R_3-2R_2} t_{\sigma-}^{(3)} t_{--}^{(2)} \nonumber\\ 
&\times &(\ket{3,\sigma}_i \ket{1,-}_{i+1} + \ket{1,-}_i 
\ket{3,\sigma}_{i+1}) \ket{3,-}_{j\neq i, j\neq i+1} \nonumber \\
&\times& \prod_{l \neq i, l \neq i+1, l \neq j} \ket{2,-}_l , \nonumber \\
\ket{\psi_{d}^{(1)}(2)} &=&\frac{1}{\sqrt{N}}\sum_{i, 
{\sigma=\pm}}\frac{-2J}{\Delta-R_2-R_3-(\sigma)R_4} t_{\sigma-}^{(4)} 
t_{--}^{(2)}\nonumber\\ 
&\times& (\ket{4,\sigma}_i \ket{1,-}_{i+1}+\ket{1,-}_i\ket{4,\sigma}_{i+1}) 
\nonumber \\
&\times& \prod^N_{l\neq i, l\neq i+1} \ket{2,-}_l , \nonumber\\ 
\ket{\psi_{d}^{(1)}(3)} &=&\frac{1}{\sqrt{N}}\sum_i \biggl[ \,
\sum_{\nu=\pm}(\ket{3,+}_i 
\ket{2,\nu}_{i+1}+\ket{2,\nu}_i \ket{3,+}_{i+1})\nonumber\\
&\times& \frac{2J t_{+-}^{(3)}t_{-\nu}^{(3)}}{2R_2+R_3+\nu R_3} \nonumber \\
& +& (\ket{3,-}_i \ket{2,+}_{i+1} + \ket{2,+}_i \ket{3,-}_{i+1} 
) \frac{Jt_{--}^{(3)}t_{-+}^{(3)}}{R_2}\biggr]\nonumber \\
&\times& \prod^N_{l \neq i, l \neq i+1} \ket{2,-}_l .
\end{eqnarray}

The equation $E_M=E_d$ leads to the upper boundary Eq.~(\ref{upperbdry})
of the $n=2$ Mott phase.  We mention that we have
also calculated the ground state energy $E_{2d}$ of the $n=2$ Mott
state doped by two excitons and find that the energy at the boundary
determined by $E_M=E_d$ is lower than that at the line determined by
$E_M=E_{2d}$. The former also matches the DMRG and the SGF results
well.

To find the lower boundary of the $n=2$ Mott lobe, we calculate the ground 
energy of the state with two excitons (holons) removed from (added to) the 
$n=2$ Mott state:
\begin{eqnarray}
E_{2h}=E_{2h}^{(0)}+E_{2h}^{(1)}+E_{2h}^{(2)},
\end{eqnarray}
with
\begin{eqnarray}
E_{2h}^{(0)} &=& \bra{\psi_{2h}^{(0)}} H_0 \ket{ \psi_{2h}^{(0)}} 
= (N-1)(2\omega+\frac{\Delta}{2}-\frac{R_2}{2}) , \nonumber \\
E_{2h}^{(1)} &=& \bra{\psi_{2h}^{(0)}} H_1\ket{\psi_{2h}^{(0)}}=0, \nonumber \\
E_{2h}^{(2)} &=& \bra{\psi_{2h}^{(1)}} H_1\ket{\psi_{2h}^{(0)}} \nonumber \\
&=&4J^2\Biggl[ (N-2) (\frac{t_{--}^{(3)2} t_{--}^{(2)2}}{\Delta+R_3-2R_2} 
+\frac{t_{+-}^{(3)2} t_{--}^{(2)2}}{\Delta-R_3-2R_2} ) \nonumber \\
&+&\frac{2 t_{--}^{(2)2}}{\Delta-R_2}\Biggr],
\end{eqnarray}
where the wave functions are 
\begin{eqnarray}
\ket{\psi_{2h}^{(0)}} &=& \frac{1}{\sqrt{N}} \sum_i\ket{0,-}_i \prod^N_{l\neq 
i}\ket{2,-}_l , \\ 
\ket{\psi_{2h}^{(1)}} &=& \frac{1}{\sqrt{N}}\sum_i\sum_{j\neq i, j\neq 
i+1}\sum_{\sigma=\pm} \Biggl[ \frac{-2J}{\Delta-2R_2-\sigma R_3} 
t_{\sigma-}^{(3)}t_{--}^{(2)}\nonumber \\ 
&\times&(\ket{3,\sigma}_i \ket{1,-}_{i+1} + \ket{1,-}_i \ket{3,\sigma}_{i+1}) 
\ket{0,-}_j  \nonumber \\ 
&\times& \prod^N_{l\neq i,l\neq i+1,l\neq j} \ket{2,-}_l  \nonumber \\
&+& \frac{-4Jt_{--}^{(2)} }{\Delta-R_2} \sum_i \ket{1,-}_i \ket{1,-}_{i+1} 
\prod^N_{l\neq i,l\neq i+1} \ket{2,-}_l \Biggr].\nonumber 
\end{eqnarray}

The equation $E_M=E_{2h}$ leads to the lower boundary, Eq.~(\ref{lowerbdry}) 
of the $n=2$ Mott phase. Note that for the lower boundary, it is necessary to 
remove two excitons to obtain the lowest energy state whereas for the upper 
boundary, we add only one exciton. This is confirmed by numerical calculations 
and leads to consequences discussed in the text.

To find the boundary of the vacuum, we need to find the energy of the
state with only 2 excitons.
\begin{eqnarray}
E_{2e}=E_{2e}^{(0)}+E_{2e}^{(1)}+E_{2e}^{(2)},
\end{eqnarray}
with
\begin{eqnarray}
E_{2e}^{(0)} &=& \bra{\psi_{2e}^{(0)}} H_0 \ket{ \psi_{2e}^{(0)}} = 
2\omega+\frac{\Delta}{2}-\frac{R_2}{2}, \nonumber \\ 
E_{2e}^{(1)} &=& \bra{\psi_{2e}^{(0)}} H_1 \ket{ \psi_{2e}^{(0)}} = 0, 
\nonumber\\
E_{2e}^{(2)} &=& \bra{\psi_{2e}^{(1)}} H_1 \ket{ \psi_{2e}^{(0)}} 
=\frac{8J^2}{\Delta-R_2} t_{--}^{(2)2}, 
\end{eqnarray}
where the wave functions are
\begin{eqnarray}
\ket{\psi_{2e}^{(0)}} &=&\frac{1}{\sqrt{N}}\sum_i \ket{2,-}_i\prod^N_{l\neq i} 
\ket{0,-}_l,   \\ 
\ket{\psi_{2e}^{(1)}} &=& \frac{-1}{\sqrt{N}}\frac{4J}{\Delta-R_2} \sum_i 
t_{--}^{(2)} \ket{1,-}_i \ket{1,-}_{i+1} \prod^N_{l\neq i, l\neq i+1} 
\ket{0,-}_l.  \nonumber
\end{eqnarray}
The vacuum boundary, Eq.~(\ref{vacbdry}), is given by the equation $E_{2e}=0$.

%%%%%%%%%%%%%%%%%%%%%%%%%%%%%%%%%%%%%%%%%%%%%%%%%%%%%%%

\end{document}